\documentclass[aps,prl,groupedaddress,draft,showpacs]{revtex4}
\begin{document}

%\hoffset-1cm

% Yale printer values
%\voffset1.5cm
%\draft
%\preprint{nucl-th/99????}
\title{$U_A(1)$ Anomaly at high temperature: the scalar-pseudoscalar splitting in QCD}
\author{Gerald~V.~Dunne and  Alex~Kovner}
\affiliation{ Physics Department, University of Connecticut, Storrs, CT 06269-3046, USA}

%\date{\today}
\begin{abstract}
We estimate the splitting between the spatial correlation lengths in the scalar and pseudoscalar channels in QCD at high temperature. The splitting is due to the contribution of the instanton/anti-instanton chains in the thermal ensemble, even though instanton contributions to thermodynamic quantities are suppressed. 
%As expected, 
The splitting vanishes at asymptotically high temperatures as $\Delta M/M\propto (\Lambda_{QCD}/T)^b$, where $b$ is the beta function coefficient. 
\end{abstract}

 
 \maketitle

%%%%%%%%%%%%%%%%%%%%%%%%%%%%%%%%%%%%%%%%%%%%%%%%%%%%%%%%%%%%%%%%%%%%%%%%%%%%%%%%%%%%%%%%
There has been a lot of interest lately in the high temperature phase of Quantum Chromodynamics \cite{Stephanov:2004wx}. Although the main thrust of the recent activity, triggered by the RHIC data, has been at temperatures close to the transition temperatures, there still are open questions about the high temperature deconfined phase of QCD. This regime may soon be experimentally accessible;  the recently published RHIC data \cite{Adare:2008fqa} suggests a temperature in the range $300 \leq T\leq 600\,{\rm MeV}$, considerably above the expected transition temperature of $T_c\approx 170$ MeV.
At asymptotically high temperatures the basic QCD physics is perturbative, however some aspects of it cannot be understood without nonperturbative effects. The generation of magnetic mass is an example of such an effect \cite{Nakamura:2003pu}. Another, simpler, set of questions has to do with contribution of instantons at high temperature \cite{Gross:1980br}. Although the instanton contribution to thermodynamic quantities like pressure and energy is negligible, they must give the leading contribution to certain correlation functions. In particular, the question of whether and when the axial $U_A(1)$ symmetry of QCD is restored above the phase transition has been debated for some time. Clearly, since the anomaly is an operator equation, it remains true on the operator level also at nonzero temperature \cite{Itoyama:1982up}. At low temperatures, the relation between the anomaly equation and $\pi^0$ decay has been elucidated in \cite{Pisarski:1997bq}. 
At very high temperatures, the effects of the anomaly are however generally believed to be very small because of increasing irrelevance of instanton contributions at high temperature \cite{Gross:1980br}. A convenient measure of the explicit breaking of the $U_A(1)$ symmetry in the theory with two massless flavors is the difference of the spatial correlation lengths in the scalar and pseudoscalar channels. This quantity has been calculated on the lattice and has been found to be small and probably nonvanishing \cite{Bernard:1996iz,Chandrasekharan:1998yx,Laermann:2003cv,Gavai:2008yv}, although the lattice studies are rather inconclusive because  the signal is numerically delicate, especially in the presence of chiral zero modes. We are, moreover, unaware of an analytic calculation of this splitting. The aim of this short note is to provide such a calculation and to demonstrate that the instanton configurations contribute decisively to this quantity even though their contribution to pressure etc. may be negligible. This work is inspired by \cite{Kogan:2001rn} where a very similar calculation was performed in the three dimensional Georgi-Glashow model, which also exhibits a deconfinement phase transition \cite{Dunne:2000vp}. Indeed there is a very close analogy between the symmetry 
structure of the 3D Georgi-Glashow model and QCD with massless quarks, as discussed in detail in \cite{Kogan:2001ex}.

We are interested in calculating the equilibrium (equal time) correlation functions
\begin{equation}
\langle S(x) S(y)\rangle\quad , \quad \langle P(x)P(y)\rangle
\end{equation}
where
\begin{equation}
S(x)=\bar\psi_i\psi_i=\bar\psi^L_i\psi^R_i+\bar\psi^R_i\psi^L_i
\quad ; \quad
P(x)=\bar\psi_i\, i\gamma_5\psi_i=i\left(\bar\psi^L_i\psi^R_i-\bar\psi^R_i\psi^L_i\right)
\end{equation}
The flavor index $i$ takes values $1,2$.
The action of the axial $U_A(1)$ symmetry as usual is
\begin{equation}
\psi^{L}_i\rightarrow e^{i\alpha}\psi^L_i\quad ; \quad
\psi^R_i\rightarrow e^{-i\alpha}\psi^R_i
\end{equation}
If the axial anomaly is absent, the diagonal correlators must vanish in the axially symmetric thermal ensemble
\begin{equation}
\langle \bar\psi^L_i(x)\psi^R_i(x)\  \bar\psi^L_i(y)\psi^R_i(y)\rangle=0=\langle \bar\psi^R_i(x)\psi^L_i(x)\  \bar\psi^R_i(y)\psi^L_i(y)\rangle
\end{equation}
and therefore
\begin{equation}
\langle S(x) S(y)\rangle = \langle P(x)P(y)\rangle
\end{equation}

Although the axial symmetry is anomalous, the effects of the anomaly  are "transmitted" to the QCD spectrum via the nonperturbative instanton and anti-instanton contributions \cite{'tHooft:1976fv,Jackiw:1977yn,Vainshtein:1981wh,Schafer:1996wv}.
At zero temperature instantons are finite action solutions of Euclidean equations of motion. In the regular gauge the vector potential of an instanton in the $SU(2)$ gauge theory is 
\begin{equation}
A_\mu^a(x)={2\eta_{a\mu\nu}x_\nu\over x^2+\rho^2}
\end{equation}
where the totally antisymmetric 't Hooft symbol is given by $\eta_{a\mu\nu}=\epsilon_{a\mu\nu}; \ \mu,\ \nu=1,2,3; \ \eta_{a\mu 4}=\delta_{a\mu}$.
The instanton in the SU(N) theory is essentially the same configuration embedded into the SU(N) group. All color orientations of the instanton have to be summed over in the path integral.  

Due to conformal symmetry of the classical QCD action, the instanton solution is characterized by a size parameter $\rho$. The action of an instanton is independent of $\rho$ classically 
\begin{equation}
S_I={8\pi^2\over g^2}
\end{equation}
The most important quantum  correction to this result stems from the fact that the coupling constant runs with scale, and so semiclassically
\begin{equation}
S_I={8\pi^2\over g^2(\rho)}
\end{equation}
The measure of integration over the instanton size is $dn\propto {d\rho\over \rho^5}(\rho\Lambda)^b$, where $b=(11N_c-2N_f)/3$ is the coefficient of the one loop $\beta$-function. The measure is peaked towards  large size instantons, which makes instanton calculations at zero temperature uncontrollable. At zero temperature therefore, although we understand qualitatively that the mass splitting between the scalars and pseudoscalars is due to instanton-like fluctuations in the vacuum, a first-principles controllable calculation is not possible. The best one can do at present is the semi phenomenological instanton liquid model \cite{Diakonov:1983hh,Shuryak:1997vd}, which  works quite well.

At finite temperature the situation is similar, as long as the temperature does not significantly exceed the nonperturbative QCD scale $\Lambda_{\rm QCD}$ \cite{Schafer:1996wv}. However, at high temperature the instanton calculations are much more under control since the temperature provides an effective infrared cutoff on the instanton size \cite{Affleck:1979gy}.
In the imaginary time formalism at finite temperature instantons are periodic solutions of the classical equations of motion with period $\beta=1/T$ in the imaginary time direction. Instantons with  core size $\rho\ll \beta$ look essentially the same as at zero temperature, but large instantons with $\rho\gg\beta$ look very different. The core of size $\beta$ is accompanied by a dyon-like field in the spatial dimensions, and so at large distances a large size instanton looks like a monopole \cite{Gross:1980br}. Most importantly, since the nonzero temperature $T$  provides an external scale, already classically the instanton action depends on the instanton size. As a result, large instantons with sizes $\rho>\beta$ are exponentially suppressed. Since the contribution to the path integral due to very small instantons is suppressed by the zero temperature measure, this means that the main contribution to physical observables comes from the instantons with the size $\rho\sim \beta$. 

An instanton at zero and finite temperature is accompanied by fermionic zero modes. There is one left-handed zero mode for  $\psi_i$ of each flavor and one right-handed zero mode for $\bar\psi_i$ of each flavor. Due to the presence of these zero modes every tunneling event associated with the instanton is accompanied by the change of the axial $U_A(1)$ charge in the vacuum, giving a concrete manifestation of the anomaly. The presence of the zero modes also modifies significantly the interaction between the instantons and anti-instantons. In  pure Yang-Mills theory, an instanton (I) and an anti-instanton (A) interact weakly with the interaction  ``potential'' decreasing as a power of the distance. Put differently, a field configuration corresponding to an Instanton/Anti-instanton pair separated by a distance $R$ (assuming the sizes of the instantons are much smaller than the separation) has the action
\begin{equation}
S_{IA}(R)\sim {16\pi^2\over g^2}\left[1+c{\rho_I^2\rho_A^2\over R^4}\right]
\end{equation}
The actual formula is somewhat more complicated since the instantons also posess a color orientation, and thus the interaction depends on the relative color orientation of the instanton and an anti-instanton. We disregard this subtlety since at high temperature the favored orientation is such that I and A are parallel in the color space \cite{Schafer:1996wv,Shuryak:1997vd}.

In the presence of the fermion zero modes the I-A interaction becomes long range. At zero temperature the wave function of the zero mode decreases as $1/R^3$ away from the center of the instanton, and as a result the I-A interaction is logarithmic. At high temperatures
on the other hand, the fermionic zero mode wave function has a characteristic exponential decay in space with the "`mass"' equal to the lowest Matsubara frequency $\pi T$. As a result the I-A interaction potential becomes linear. For an I-A pair separated by a large distance R in the spatial direction the interaction potential at high temperature due to $2N_f$ zero modes is
\begin{equation}
S_{IA}(R)\approx S_I+S_A+ 2N_f\pi TR
\end{equation}
At these high temperatures the instantons are like monopoles, and the ``confining''  linear potential between a monopole and an anti-monopole binds the monopoles into pairs of net zero topological charge. It is believed that the transition to the ``molecular''  phase coincides with the chiral symmetry restoration and the deconfinement phase transition and happens around $T_c \approx 170$ Mev in the theory with two massless flavors \cite{Schafer:1996wv}.

A simpler model which exhibits a very similar behavior and instanton binding above the deconfinement phase transition is the Georgi-Glashow model in 2+1 dimensions \cite{Dunne:2000vp,Kogan:2001rn,Kogan:2001ex}. Although in this model the instantons are bound in pairs at high temperature, they nevertheless have a direct effect on some observables. As was shown in \cite{Kogan:2001rn}, in the 2+1 dimensional Georgi-Glashow model they give the main contribution to the mass (spatial correlation length) splitting between the scalar and the pseudoscalar channels. We expect similar effects in QCD.  Instanton effects should be there in a theory with any number of flavors, but they are simplest to see in the two flavor case, so in the following we therefore concentrate on $N_f=2$. 

Consider the equal time correlation function 
\begin{equation}
G_{LL}(x-y)=\langle \bar\psi^R_i(x)\psi^L_i(x)\  \bar\psi^R_i(y)\psi^L_i(y)\rangle
\end{equation}
As mentioned above, in the absense of the $U_A(1)$ anomaly this correlation function vanishes, since it has a net axial $U_A(1)$ charge. Thus perturbative calculation of this correlator at high temeperature gives a vanishing result. It is however easy to see that this correlator does not vanish on a single Instanton configuration. The instanton has four zero modes -- exactly the right number to be saturated by the fermionic operators in the observables. Thus, as long the points $x$ and $y$ are very far away from each other, $|\vec x-\vec y|\ll 1/T$, an instanton centered at a point $\vec a$  gives a contribution 
\begin{equation}\label{onei}
\langle \bar\psi^L_i(x)\psi^R_i(x)\ \bar\psi^L_i(y)\psi^R_i(y)\rangle_I\approx aT^6 e^{-S_I}e^{-2\pi T\{|\vec x-\vec a|+|\vec y-\vec a|\}}
\end{equation}
where $a$ is a constant of order one.
The maximal contribution comes from the instantons located along the straight line connecting points $x$ and $y$. The instanton contribution does not depend on the position of the instanton along the line. Integrating over the position of the instanton with the weight $Tdz$ we obtain
\begin{equation}
\langle \bar\psi^R_i(x)\psi^L_i(x)\ \bar\psi^R_i(y)\psi^L_i(y)\rangle_I\approx (aT^6) (cT)|\vec x-\vec y| e^{-S_I(\beta)}e^{-2\pi T |\vec x-\vec y |}
\end{equation}
where we have set the instanton size to be equal to the inverse temperature and have allowed for a dimensionless constant $c$, of order one, which must appear due to the integration over the instanton sizes. It should be possible to estimate $c$ using the dependence of the instanton action on $\rho$ \cite{Gross:1980br}, but our main interest in this note is to establish  the parametric dependence of the correlation length on temperature.
The next contribution comes from a configuration of two instantons and one anti-instanton, alternating along the straight line connecting $\vec x$ and $\vec y$. The zero modes are again completely saturated and so this contribution does not vanish. The main contribution comes from the configuration where two of the zero modes of each instanton are saturated by the zero modes of the anti-instanton. The typical distances between the instantons clearly is greater than their size, and thus we can neglect all interactions between them except for that induced by the fermionic zero modes. The contribution of this configuration gives
\begin{equation}
\langle \bar\psi^R_i(x)\psi^L_i(x)\  \bar\psi^R_i(y)\psi^L_i(y)\rangle_{IAI}\approx (aT^6) {1\over 2}c^3T^3|\vec x-\vec y|^3 e^{-3S_I(\beta)}e^{-2\pi T|\vec x-\vec y|}
\end{equation}
The leading contribution from a configuration with an alternating  chain of $n$ instantons and $n-1$ anti-instantons is
\begin{equation}
\langle \bar\psi^R_i(x)\psi^L_i(x)\  \bar\psi^R_i(y)\psi^L_i(y)\rangle_{I^nA^{n-1}}\approx 
(aT^6)(cTe^{-S_I(\beta)})^{2n-1}e^{-2\pi T|\vec x-\vec y|}
\int_x^ydy_n\int_x^{y_n}dx_{n-1}\int_x^{x_{n-1}}dy_{n-1}...
\int_x^{x_1}dy_1
\end{equation}
Summing all the contributions we obtain 
\begin{equation}\label{ll}
\langle \bar\psi^R_i(x)\psi^L_i(x)\ \bar\psi^R_i(y)\psi^L_i(y)\rangle\approx (aT^6)e^{-2\pi T|\vec x-\vec y|}\sinh \left(cTe^{-S_I(\beta)}|\vec x-\vec y|\right)
\end{equation}
As already mentioned,  we have assumed that all the instantons and anti-instantons have the same orientation in color space, because at high T
%Strictly speaking one of course has to integrated over the relative color orientation. It 
%is known however  that 
the integration over the orientations is dominated by the configuration where instantons and anti-instantons are parallel \cite{Schafer:1996wv}. The integration over orientations will only affect the value of the pre-exponential factor $c$.

Turning our attention to the axially symmetric correlator, we note that perturbatively
\begin{equation}
\langle \bar\psi^L_i(x)\psi^R_i(x)\  \bar\psi^R_i(y)\psi^L_i(y)\rangle_{\rm perturbative}\approx
(aT^6)\, e^{-2\pi T|\vec x-\vec y|}
\end{equation}
The origin of the exponential factor is now not the zero modes but rather just the contribution of the free lowest Matsubara frequency contribution. The constant factor $aT^6$ is the same as in eq.(\ref{onei}) etc. since it originates from the disconnected piece associated with points $x$ and $y$ separately. Clearly instanton/anti-instanton chains give a contribution to this correlator in exactly the same way as before. The only difference is that now the number of instantons and anti-instantons in the chain must be equal in order to saturate all the fermionic zero modes. The reult is then
\begin{equation}\label{lr}
\langle \bar\psi^L_i(x)\psi^R_i(x)\  \bar\psi^R_i(y)\psi^L_i(y)\rangle\approx (aT^6)e^{-2\pi T|\vec x-\vec y|}\cosh \left(cTe^{-S_I}|\vec x-\vec y|\right)
\end{equation}

From equations (\ref{ll}) and (\ref{lr}) we find 
\begin{equation}
\langle S(x)S(y)\rangle\propto e^{-M_S|x-y|}\quad ; \quad \langle P(x)P(y)\rangle\propto e^{-M_P|x-y|}\quad ; \quad  \langle S(x)P(y)\rangle=0
\end{equation}
with
\begin{equation}
M_S=2\pi T-c\, Te^{-S_I(\beta)}; \ \ \ \ M_P=2\pi T+c\, Te^{-S_I(\beta)}
\end{equation}
Thus we find that the splitting between the scalar and pseudoscalar correlation lengths is given by
\begin{equation}
{\Delta M\over M}\propto e^{-S_I(\beta)}\propto \left({\Lambda_{QCD}\over T}\right)^b
\end{equation}
which is the advertised result. In the end, this is essentially the expected semiclassical quantum mechanical splitting determined by exponentiation of the single-instanton action \cite{Landau,Coleman:1978ae}. The key observation is that at high $T$  the problem reduces to one of a  {\it linear} chain of alternating instantons and anti-instantons, with the total number being either different by one (anti-symmetric correlator), or equal  (symmetric correlator). This is very similar to the behavior found in the Georgi-Glashow model \cite{Kogan:2001rn}, as we expected.

We note that although we have discussed isosinglet correlators, the results apply also to isovectors since the chiral $SU(2)\otimes SU(2)$ symmetry is restored in the high temperature phase. Thus, for example we also have
\begin{equation}
{M_{a_0}-M_\pi\over M_\pi}\propto\left({\Lambda_{QCD}\over T}\right)^b
\end{equation}
Although this ratio is not parametrically suppressed at $T\sim 2T_c$, the large power $b\sim 10$ may explain the fact that at these temperatures the difference in the correlation lengths is difficult to detect in lattice calculations \cite{Gavai:2008yv}. It would be very interesting if the high $T$ regime could be probed further with high statistics lattice computations. 

A simple consistency check on our calculation is to add a $\theta$ term to the QCD Lagrangian. In the chiral limit the spectrum of masses and correlation lengths should not depend on the value of $\theta$ as it can be eliminated by the anomalous $U(1)$ rotation. This means that with $\theta\ne 0$ we should recover the same correlation lengths, but in the correlators of the axially rotated operators. We can include the effect of the QCD $\theta$ angle in our calculation by noting that the correlators with one extra instanton or anti-instanton in the chain will acquire a phase, while the correlators with equal numbers of instantons and anti-instantons will be unchanged. Thus, 
$G_{LL}$ acquires an extra phase $e^{i\theta}$, $G_{RR}$ acquires an extra phase $e^{-i\theta}$, while $G_{LR}$ and $G_{RL}$ are unchanged. Then we find
\begin{eqnarray}
\langle S(x)S(y)\rangle\propto 2e^{-M|x-y|}\left[\cosh(\Delta M|x-y|)-\cos\theta\,\sinh(\Delta M|x-y|)\right]\nonumber\\
\langle P(x)P(y)\rangle\propto 2e^{-M|x-y|}\left[\cosh(\Delta M|x-y|)+\cos\theta\,\sinh(\Delta M|x-y|)\right]\nonumber\\
\langle P(x)S(y)\rangle=\langle S(x)P(y)\rangle\propto 2e^{-M|x-y|}\sin\theta\,\sinh(\Delta M|x-y|)
\end{eqnarray}
with $M={1\over 2}(M_S+M_P), \ \ \Delta M={1\over 2}(M_P-M_S)$.
The eigenvalues of the correlator matrix are the same as before while the eigenvectors are rotated precisely by the axial rotation, as expected.
%The screening masses do not change, but there is mixing between states, characterized by the theta angle. The point $\theta=2\pi$ is equivalent to $\theta=0$,
%while at $\theta=\pi$ we do not change the spectrum, but rename the states, effectively interchanging $S$ and $P$.

We conclude with two brief comments. First, for a theory with more than two massless flavors, the I-A chains do not contribute to the correlation function of two fermionic bilinears, since the number of zero modes on an instanton does not match the number of fermionic operators. Instead they contribute to the Green's functions with $N_f$ bilinears.  
Second, at high temperature in QCD the instantons can be thought of as Skyrmions of the field $A_0^a$ in the dimensionally reduced theory \cite{Atiyah:1989dq}. This is very similar to the situation in the 3D Georgi-Glashow model, where instanton-monopoles became vortices of the abelian part of $A_0$. The interesting difference is that it is easy to understand the  linear interaction between vortices \cite{Kogan:2001rn,Kovner:2002yv} as a direct result of the nonexistence of a continuous $U(1)$ symmetry. On the other hand it is more difficult to understand a linear potential between Skyrmions, since the Skyrmion field decays fast at infinity. Since this linear interaction is the consequence of the fermionic zero modes, it would be interesting to understand how this interaction arises in the effective Lagranian of the dimensionally reduced theory due to the integration of the quark fields.

\bigskip
We  acknowledge support through the DOE  grant DE-FG02-92ER40716.

\end{document}